\documentclass{emulateapj}
\usepackage{apjfonts}
\usepackage{epsf}
\begin{document}

\slugcomment{}
\shortauthors{J. M. Miller et al.}
\shorttitle{Patchy Disks in ULXs}

\title{Patchy Accretion Disks in Ultraluminous X-ray Sources}

\author{J.~M.~Miller\altaffilmark{1}, M. Bachetti\altaffilmark{2,3},
  D. Barret\altaffilmark{2,3}, F. A. Harrison\altaffilmark{4},
  A. C. Fabian\altaffilmark{5}, N. A. Webb\altaffilmark{2,3},
  D. J. Walton\altaffilmark{4}, V. Rana\altaffilmark{4}}
 
\altaffiltext{1}{Department of Astronomy, University of Michigan, 500
Church Street, Ann Arbor, MI 48109-1042, USA, jonmm@umich.edu}

\altaffiltext{2}{Universite de Toulouse, UPS-OMP, IRAP, Toulouse, France}

\altaffiltext{3}{CNRS, Intitut de Recherche in Astrophysique et
  Planetologie, 9 Av. Colonel Roche, BP 44346, F-31028, Toulouse Cedex
  4, France}

\altaffiltext{4}{Cahill Center for Astronomy \& Astrophysics, California Institute of Technology, Pasadena, CA 91125, USA}

\altaffiltext{5}{Institute of Astronomy, University of Cambridge,
Madingley Road, Cambridge, CB3 OHA, UK}


\label{firstpage}

\begin{abstract}
The X-ray spectra of the most extreme ultra-luminous X-ray sources --
those with $L \geq 10^{40}~ {\rm erg}~ {\rm s}^{-1}$ -- remain
something of a mystery.  Spectral roll-over in the 5--10~keV band was
originally detected in in the deepest {\it XMM-Newton} observations of
the brightest sources; this is confirmed in subsequent {\it NuSTAR}
spectra.  This emission can be modeled via Comptonization, but with
low electron temperatures ($kT_e \simeq$2~keV) and high optical depths
($\tau \simeq 10$) that pose numerous difficulties.  Moreover,
evidence of cooler thermal emission that can be fit with thin disk
models persists, even in fits to joint {\it XMM-Newton} and {\it
  NuSTAR} observations.  Using NGC 1313 X-1 as a test case, we show
that a patchy disk with a multiple temperature profile may provide an
excellent description of such spectra.  In principle, a number of
patches within a cool disk might emit over a range of temperatures,
but the data only require a two-temperature profile plus standard
Comptonization, or three distinct blackbody components.  A mechanism
such as the photon bubble instability may naturally give rise to a
patchy disk profile, and could give rise to super-Eddington
luminosities.  It is possible, then, that a patchy disk (rather than a
disk with a standard single-temperature profile) might be a hallmark
of accretion disks close to or above the Eddington limit.  We discuss
further tests of this picture, and potential implications for sources
such as narrow-line Seyfert-1 galaxies (NLSy1s) and other low-mass
active galactic nuclei (AGN).
\end{abstract}

\section{Introduction}
Ultra-luminous X-ray (ULXs) are accretion-powered sources that appear
to violate the isotropic Eddington limit for a fiducial $M =
10~M_{\odot}$ black hole.  In practice, a better lower limit is $L_{X}
\geq 2\times 10^{39}~{\rm erg}~ {\rm s}^{-1}$ (Irwin, Athey, \&
Bregman 2003).  Most or all of the sources that barely qualify as ULXs
are likely to be stellar-mass black holes similar to those known in
the Milky Way.  The set of ULXs with $L_{x} \geq 10^{40}~ {\rm erg}~ {\rm
  s}^{-1}$ is far more interesting, as these sources more strongly
indicate super-Eddington accretion or elevated black hole masses.

Only a small number of these extreme ULXs are found at distances that
permit sensitive spectra in reasonable observation times (note that
Sutton et al.\ 2012 and Gladstone 2013 define ``extreme''
differently).  Yet, detailed studies have revealed multiple components
in the spectra of these ULXs, including soft disk--like components.
The low temperature values typical of these components ($kT \simeq
0.2$~keV) may indicate accretion onto more massive black holes
(e.g. $10^{2-3}~ M_{\odot}$; Miller et al.\ 2003, 2004, 2013).  Mass
estimates depend strongly on numerous assumptions (e.g. Soria 2011,
Miller et al.\ 2013), not least the idea that black holes in ULXs
accrete in a mode that is observed in sub-Eddington stellar-mass black
holes and in AGN.

However, ULXs may show spectral phenomena that stellar-mass black
holes and AGN may not.  In particular, deep {\it XMM-Newton} spectra
show evidence of a roll-over in the 5--10~keV band.  This peculiar
hard component can also be modeled as Comptonization of low
temperature photons (e.g. Stobbart et al.\ 2006, Gladstone et
al.\ 2009).  In constrast to the hot, optically--thin Comptonization
regions inferred in other black holes, the roll-over in ULXs requires
a cool, optically--thick region ($kT_e \simeq 2$~keV; $\tau \simeq
10$).  Unless the corona is heated locally and/or powered magnetically
-- in some manner that would prevent substantial heating -- it is easy
to show that such a cool corona must be huge in order to generate the
bulk of the observed $L_{X} \geq 10^{40}~ {\rm erg}~ {\rm s}^{-1}$
that is observed (see, e.g., Merloni \& Fabian 2001 concerning thermal
and magnetic coronae).  Not only is it difficult to envisage a very
large region that can be characterized by a single temperature, but
the outer part of the putative corona may only be marginally bound to
the black hole.

Some extreme ULXs have been detected up to 40~keV in recent {\it
  NuSTAR} (Harrison et al.\ 2013) observations, and the roll--over
found in prior 0.3--10.0~keV spectra is strongly confirmed (e.g Walton
et al.\ 2013a, 2014; Bachetti et al.\ 2013; Rana et al.\ 2014).
Low--temperature Comptonization again provides acceptable fits, but
again is not a unique description, and the problems of such coronae
remain.  

Putative Comptonization components with such low electron temperatures
and high optical depths, are fairly similar to a blackbody
(e.g. Stobbart et al.\ 2006; Kajava et al.\ 2012; Middleton et
al.\ 2011).  Miller et al.\ (2013) recently reported positive
correlations between the temperature and luminosity of the cool ($kT
\simeq 0.2$~keV) disk--like components in ULXs, suggesting that a disk
interpretation may be also be correct or that component (less
stringent data selection and modeling previously led to different
conclusions; see Kajava \& Poutanen 2009; Pintore \& Zampieri 2012).
In view of that result -- and the blackbody--like nature of the harder
component that rolls-over at high energy -- Miller et al.\ (2013)
suggested that both components may originate in the disk, and that ULX
spectra may be consistent with ``patchy'' or inhomogeneous disks (see
Dexter \& Quataert 2012; also see Begelman 2002, Dotan \& Shaviv
2011).  This Letter explicitly examines the possibility of
inhomogeneous disks in extreme ULXs.

\section{Data Reduction}
NGC 1313 X-1 was selected as our test case because it has been
observed on numerous occasions with {\it XMM-Newton}, and on two
occasions with both {\it XMM-Newton} and {\it NuSTAR}.  The data
considered in this work are exactly the data from two epochs of
observations treated in Bachetti et al.\ (2013), and the reader is
referred to that paper for details of the observations and data reduction.

Again following Bachetti et al.\ (2013), we fit the {\it
  XMM-Newton}/EPIC-pn and EPIC-MOS2 spectra in the canonical
0.3--10.0~keV band, and the {\it NuSTAR}/FPMA and FPMB spectra in the
3--30~keV band.  However, whereas Bachetti et al.\ (2013) adopted
different grouping schemes for the {\it NuSTAR} data depending on the
behaviors of specific models, we binned every spectrum to require at
least 20 counts per bin for every model (in order to ensure the
validity of $\chi^{2}$ statistics; Cash 1979).  Grouping was
accomplished using the FTOOLs suite (specifically ``grppha'') and all
spectral fits were made using XSPEC version 12.8.0 (Arnaud 1996).

The spectral fits were made allowing an overall multiplicative
constant to float between them.  All other parameters were linked
across different spectra and jointly determined by the fit.  Last, in
all fits, the absorption along the line of sight was characterized
using a single $tbabs$ model with the proper abundances and cross
sections (Wilms, Allen, \& McCray 2000).  All uncertainties in this
work are 1$\sigma$ confidence errors.

\section{Analysis and Results}
As an initial test of the viability of a patchy disk explanation for
ULX spectra and accretion flows, we fit a model consisting of two
``diskbb'' components (Mitsuda et al.\ 1984) to the spectra from both
epochs.  The hotter component may be a simple
blackbody, but if the various patches are distributed over even a
small range in disk radius, they may also have a small run in
temperature, and a disk blackbody profile is not unreasonable.

As can be seen in Figure 1, most of the observed spectra can be
accounted for with such a model, including the putative thermal,
disk--like component at low energy, and the harder component peaking
around $\simeq$5~keV and rolling-over to 10~keV.  This simple model is
even a relatively good fit, yielding $\chi^{2}/\nu = 1.07$ (see Table
1).  It is clear, however, that the two disk components fail to fit
all of the high energy flux captured with {\it NuSTAR}.  Moreover,
these fitting results are not as good as those achieved in fits to
similar data in Miller et al.\ (2013), nor as good as fits to the
exact same spectra in Bachetti et al.\ (2013).

In order to capture the additional high energy flux, we next
considered models that added different Comptonization prescriptions.
The ``compTT'' component is a physical model that allows the electron
temperature of the corona ($kT_{e}$) and its optical depth ($\tau$) to
be measured directly (Titarchuk 1994).  As noted above, it is often
fit in combination with ``diskbb'' components, even when modeling ULX
spectra.  A model consisting of two independent ``diskbb'' components,
each with a corresponding ``compTT'' component ($diskbb + compTT +
diskbb + compTT$), was therefore explored.  The seed photon
temperature in each ``compTT'' component was linked to the temperature
of its corresponding disk component, but the ``compTT'' electron
temperatures, optical depths, and flux normalizations were allowed to
float independently.

Excellent fits were obtained with this model ($\chi^{2}/\nu \simeq
1.00$), but the Comptonization parameters were largely unconstrained.
As just one example of the kind of Comptonization that might account
for the hard flux, the electron temperature and optical depth were
fixed at $kT_{e} = 100$~keV and $\tau = 0.1$, respectively.  Such
values are representative of the coronae inferred in better--understood
black holes.  Indeed, in the ``very high'' or ``steep power-law
state'' -- to which ULXs may bear some resemblance -- the electron
temperature may be even higher (Tomsick et al.\ 1999).  With this
simplification, excellent fits were again achieved (see Figures 2 and
3, and Table 1).  Importantly, the overall fit to each epoch was
improved by $\Delta \chi^{2} > 100$ relative to fits without
Comptonization (see Table 1).  This
model merely represents one simplistic description of Comptonization,
that makes contact with the coronae and Comptonizing regions implied
in more familiar Galactic black holes.  Future explorations of ULX
spectra in terms of patchy disks may need to examine different, less
idealized possibilities.

Next, we made fits with two ``diskbb'' components, both modified by
the ``simpl'' convolution function: $simpl\times (diskbb + diskbb)$.
``Simpl'' is a Comptonization model that does not provide measurements
of the electron temperature nor optical depth; rather, it merely
scatters a thermal distribution into a power-law and reports the
fraction of the incident radiation that is scattered (Steiner et
al.\ 2009).  As with the individual (but very similar) ``compTT''
components, this model achieved excellent fits (see Figures 2 and 3,
and Table 1).  The power-law index was not well constrained in our
fits, but hard indices were less favorable than relatively soft
indices.  After some experimentation, $\Gamma = 3.5$ (e.g. McClintock
\& Remillard 2006) was fixed as a representative value in our fits
(see Table 1).  This index is within the range observed in
stellar-mass Galactic black holes in the ``very high'' or ``steep
power-law'' state.

In broad terms, the relative emitting areas implied by these
phenomenological patchy disk models make sense.  Assuming the distance
to NGC 1313 is 3.7~Mpc (Tully 1988), and assuming that the disk in NGC
1313 X-1 is viewed at an inclination of 60$^{\circ}$, a
normalization of $K_{disk-1} = 16$ implies a radius of $R\simeq
2100$~km (see Table 1).   A flux
normalization of $K_{disk-2} = 4\times 10^{-3}$ is consistent with the
values of the hotter ``patches'' in Table 1.  This nominally
corresponds to an emitting area of just $R\simeq 30$~km.  

There is no a priori reason two expect a disk with patches of a single
temperature.  It may be more physically realistic to consider patches
with many different tempratures.  Therefore, we also examined how many
distinct disk blackbody or simply blackbody components ae required to
fit the data.  The results of these fits are given in the final
pairing of models in Table 1.  We find that three components are
sufficient to fit the data, with an inverse relationship between
temperature and emitting area.  (No additional hard Comptonizing
component is required.)

The temperature constrast measured in all of our simple models is
a factor of 7--9, whereas Dexter \& Quataert (2012) considered a
factor of a few.  Scattering in an atmosphere and/or within the corona
artificially hardens disk spectra (Merloni,
Ross, \& Fabian 2002).  While a color correction factor of
$T_{col}/T_{eff} = f_{col} = 1.7$ is canonical for accretion disks in
the ``high soft'' state, higher correction factors are possible, and
may be required in other states, particularly when a corona is present
(e.g. $f_{col} \geq 3$; Merloni, Fabian, \& Ross 2000).  If the hot
patches are more strongly distorted than the larger cool disk, the
true temperature difference would be consistent with the contrast
considered by Dexter \& Quataert (2012).  Implied emitting areas would
increase by $f_{col}^{2}$.

\section{Discussion and Conclusions}
Following the suggestion of patchy or inhomogeneous accretion disks in
ULXs in Miller et al.\ (2013), we have specifically examined the
ability of phenomenological descriptions of a ``patchy'' accretion
disk (e.g. Dexter \& Quataert 2012) to describe the spectra of extreme
ULXs.  At least in the case of NGC 1313 X-1, our fits to two epochs of
joint {\it XMM-Newton} and {\it NuSTAR} spectra suggest that a patchy
disk is potentially an excellent description of the observed spectra.
Importantly, such models likely avoid the need for very large but cool
Comptonization regions implied by recent spectral models
(e.g. Gladstone et al.\ 2009).

Theoretical
work has found evidence of photon bubble instabilities (Gammie 1998,
Begelman 2001, 2002) that might naturally give rise to a patchy
temperature profile and locally super-Eddington flux levels.  A patchy
disk need not be a unique signature of a photon bubble instability;
other instabilities or mechanisms might be able to produce a similar
effect.  It is possible that a two--temperature or patchy disk is the
natural signature of near-Eddington or super-Eddington accretion, and
that a compelling correspondence between theory and observations has
been identified.  But, it is important to consider caveats and means
of testing this scenario.

King et al.\ (2001) described a model for super-Eddington accretion in
ULXs wherein a funnel-like geometry develops in the inner disk,
mechanically beaming some radiation and foiling luminosity estimates
based on isotropic emission.  The hot, small region implied by our
two-temperature models could correspond to the inner wall of a funnel.
However, recent observations of the extreme ULX NGC 5408 X-1 have
revealed X-ray flux dips typical of disks viewed at {\it high}
inclination (e.g. Pasham \& Strohmayer 2013, Grise et al.\ 2013).
Moreover, recent radio observations of Holmberg II X-1 -- which can
also be fit with this spectral model -- reveal three components (Cseh
et al.\ 2014), potentially indicating a jet system viewed at high
inclination.  It is possible that the hot disk emission corresponds to
patches rather than to a line of sight that peers into a funnel.

If ULXs are not all viewed at low inclination angles, it is possible
that their radiation is nearly isotropic, and that ULXs with $L_{X}
\geq 10^{40}~ {\rm erg}~ {\rm s}^{-1}$ harbor black holes with masses
above the range known in Galactic X-ray binaries.  This may be
supported by recent evidence that cool thermal components in ULXs
appear to be broadly consistent with the $L \propto T^{4}$ trend
expected for standard thin disk accretion (Miller et al.\ 2013).
However, this does not explain why the high energy spectra of extreme
ULXs differ so markedly from stellar-mass black holes and most AGN
(Bachetti et al.\ 2013, Walton et al.\ 2013a, 2014).  A patchy disk
model may be able to account for both spectral features.  

A wider set of sources and states must be sampled in order to better
test the possibility of patchy disks in ULXs.  Variability below
10~keV might be ascribed to changes in the characteristic temperatures
-- and relative emitting areas -- of cooler and hotter disk
regions.  Such changes might be linked to the mass accretion rate
through the disk.  If a mechanism like the photon bubble instability
is at work, spectral variations might also be tied to changes in the
disk magnetic field (which could itself be linked to the mass
accretion rate).  Spectral variability above approximately 10~keV
might be driven partially by changes in an hotter disk component, but
also changes in a corona.  New observations of Holmberg
IX X-1 in different states may be consistent with a patchy accretion
disk model (Walton et al.\ 2014).  It is possible that different flux
ratios from the larger disk and hot patches could account for
differences between very soft sources (such as NGC 5408 X-1), and
harder sources (like Holmberg IX X-1).

If patchy disks power ULXs to super-Eddington luminosities, a strong
wind might be driven (but see Begelman 2001).  However, winds have not
been conclusively detected in ULXs.  Indeed, narrow absorption lines
with strengths comparable to features detected in Galactic sources are
ruled out at high confidence levels in deep spectra (Walton et
al.\ 2012, 2013b; Pasham \& Strohmayer 2012).  Emission lines from an
outflow perpendicular to the line of sight with equivalent widths
comparable to SS 433 ($few \times 100$~eV; see, e.g., Marshall et
al.\ 2013) are also ruled out.  Potential low-energy lines in NGC 5408
X-1 have been discussed as having a wind origin (Middleton et
al.\ 2014), but the lines may also be described in terms of diffuse
emission with a constant flux (e.g. Miller et al.\ 2013).

If the disk is only super-Eddington in localized regions -- the hot
patches -- then radiation pressure would only drive winds in localized
regions.  Moreover, even if the temperature profile is more shallow
than the $T \propto R^{-3/4}$ derived for standard thin disks, the
emission and wind regions will be concentrated close to the black
hole.  Any winds will therefore be highly ionized and -- assuming the
wind must retain some of the angular momentum of its launching point
-- highly smeared by rotation.  Such effects would certainly inhibit
the detection of a wind via {\it narrow} absorption lines.

It is possible that strong
QPOs could be generated via a patchy disk profile.  This might also
cause the QPOs to be stronger at the higher energies that correspond
to the hot patches; this would match observed relations in more
standard X-ray binaries.  QPOs are observed in some extreme ULXs
(e.g. M82 X-1, NGC 5408 X-1; Strohmayer \& Mushotzky 2003, Strohmayer
et al.\ 2007).  Recent
simulations suggest that patchy disk profiles do not necessarily lead
to QPO production (Armitage \& Reynolds 2003), but additional
theoretical work and observational tests may be required.

Relativistically--blurred disk reflection is the most likely
explanation for the bulk of the ``soft excess'' observed in NLSy1s
(Fabian et al.\ 2009, Kara et al.\ 2013).  Yet,
even in a case such as 1H 0707$-$495 -- wherein the disk reflection
spectrum is particularly strong and clear -- a very soft blackbody
component is still required at the edge of the band pass (Fabian et
al.\ 2009).  It is possible that this emission represents the high
energy tail of a patchy accetion disk that puts out more power in its
lower--temperature UV/EUV component.  It is interesting to note that
some sources identified by Greene \& Ho (2007) appear to have similar
X-ray spectra, and ``slim'' or super-Eddington disk models may apply
in some cases.

\vspace{0.2in}

JMM acknowledges helpful conversations with Mitch Begelman, Jason
Dexter, and Richard Mushotzky.  This work was supported under NASA
Contract No. NNG08FD60C, and made use of data from the NuSTAR mission,
a project led by the California Institute of Technology, managed by
the Jet Propulsion Laboratory, and funded by NASA.

\clearpage

\begin{table}[t]
\caption{Example Phenomenological Patchy Disk Models}
\begin{footnotesize}
\begin{center}
\begin{tabular}{lllllllllll}
\tableline
\tableline

Epoch   & $N_{H}$ & $kT_{disk-1}$ &  $K_{disk-1}$ &  $kT_{disk-2}$ &  $K_{disk-2}$ &  $kT_{e}$ &  $\tau$  &  $K_{comptt-1}$ & $K_{comptt-2}$ & $\chi^{2}/\nu$ \\      ~     &  ($10^{21}~ {\rm cm}^{-2}$) & (keV) & ~ & (keV) & ($10^{-3}$) & (keV) &  ~       &    ($10^{-6}$) &  ($10^{-7}$) &  ~   \\

\tableline

1 &  2.4(1) &  0.375(5) &  4.7(3) &   2.64(5)   &      2.8(1) &    --  &     --    &    --    &      --    &       2032/1878 \\

2 &  2.4(1) &  0.376(6) &  4.8(4) &   2.61(4)   &      3.1(2) &    --  &    --     &   --     &     --     &      2064/1935 \\

\tableline

1 &  2.73(7) &  0.27(2) &   17(4) &    2.4(2)   &       3(1)  &     100* &   0.1*  &    5(1)   &     4(2)  &       1917/1876 \\

2 &  0.265(7) & 0.27(2) &   15(3) &    2.3(2)   &       4(1)  &     100* &   0.1*  &    4(1)   &    5(2)   &      1939/1933 \\

\tableline

 ~  & $N_{H}$ & $kT_{disk-1}$ &  $K_{disk-1}$ &  $kT_{disk-2}$ &  $K_{disk-2}$ &  ~ & ~ & $\Gamma$ & $f_{scat}$  &  $\chi^{2}/\nu$ \\
    ~     &  ($10^{21}~ {\rm cm}^{-2}$) & (keV) & ~ & (keV) & ($10^{-3}$) & ~ & ~ & ~ & ~ & ~ \\ 
\tableline

1 &  2.7(1) &  0.28(1) &   16(3)  &   2.30(6) &  4.0(4) & ~ & ~  &  3.5*  &    0.42(5) &    1937/1877 \\

2 &  2.7(1) &  0.28(1) &   16(3)  &   2.23(5) &  4.6(4) & ~ & ~  &  3.5*  &    0.42(5) &    1954/1934 \\

\tableline

 ~  & $N_{H}$ & $kT_{disk-1}$ &  $K_{disk-1}$ &  $kT_{disk-2}$ &  $K_{disk-2}$ & $kT_{disk-3}$  & $K_{disk-3}$  & ~ & ~  &  $\chi^{2}/\nu$ \\
    ~     &  ($10^{21}~ {\rm cm}^{-2}$) & (keV) & ~ & (keV) & ($10^{-2}$) & (keV) & ($10^{-4}$) & ~ & ~ & ~ \\ 

\tableline

1   & 2.6(1) &  0.33(1) & 7.9(9) &  1.5(2) & 1.1(3) &  4.1(4) &  4(1) & ~ & ~ &  1923/1876 \\

2   & 2.6(1) &  0.33(1) & 7.8(9) & 1.6(2)  & 0.9(1) &  4.2(5) &  3(2) & ~ &  ~ & 1942/1933 \\

\tableline
\end{tabular}
\vspace*{\baselineskip}~\\ \end{center} 
\tablecomments{The table above lists values obtained from spectral
  fits to two epochs of joint {\it XMM-Newton} and {\it NuSTAR}
  observations of NGC 1313 X-1.  The first pairing of models consist
  of only two simple ``diskbb'' components.  The second pairing adds
  Comptonization components via ``compTT''.  More realisitic fits to
  better data might require different electron temperatures and
  optical depths for each ``compTT'' component; however, a broad range
  of combinations give acceptable fits to these spectra.  Values of
  $kT_{e} = 100$~keV and $\tau = 0.1$ were selected to demonstrate
  that standard Comptonization is compatible with the data.  The flux
  normalizations of the ``compTT'' components floated independently.
  The third pairing attempts to model potential Comptonization via
  ``simpl'', which acted on both disk components together.  The
  power-law index in the model was poorly constrained but required to
  be fairly steep.  A value of $\Gamma = 3.5$ was selected as it gives
  excellent fits and is commensurate with the ``very high'' or ``steep
  power-law'' states observed in standard black hole X-ray binaries.
  The bottom pairing consists of models with three disk blackbody
  components, with no Comptonization.  Three simple blackbody
  components give equally good fits and approximately the same
  temperatures.}
\vspace{-1.0\baselineskip}
\end{footnotesize}
\end{table}
\medskip

\clearpage

\begin{figure}
\includegraphics[scale=0.35,angle=-90]{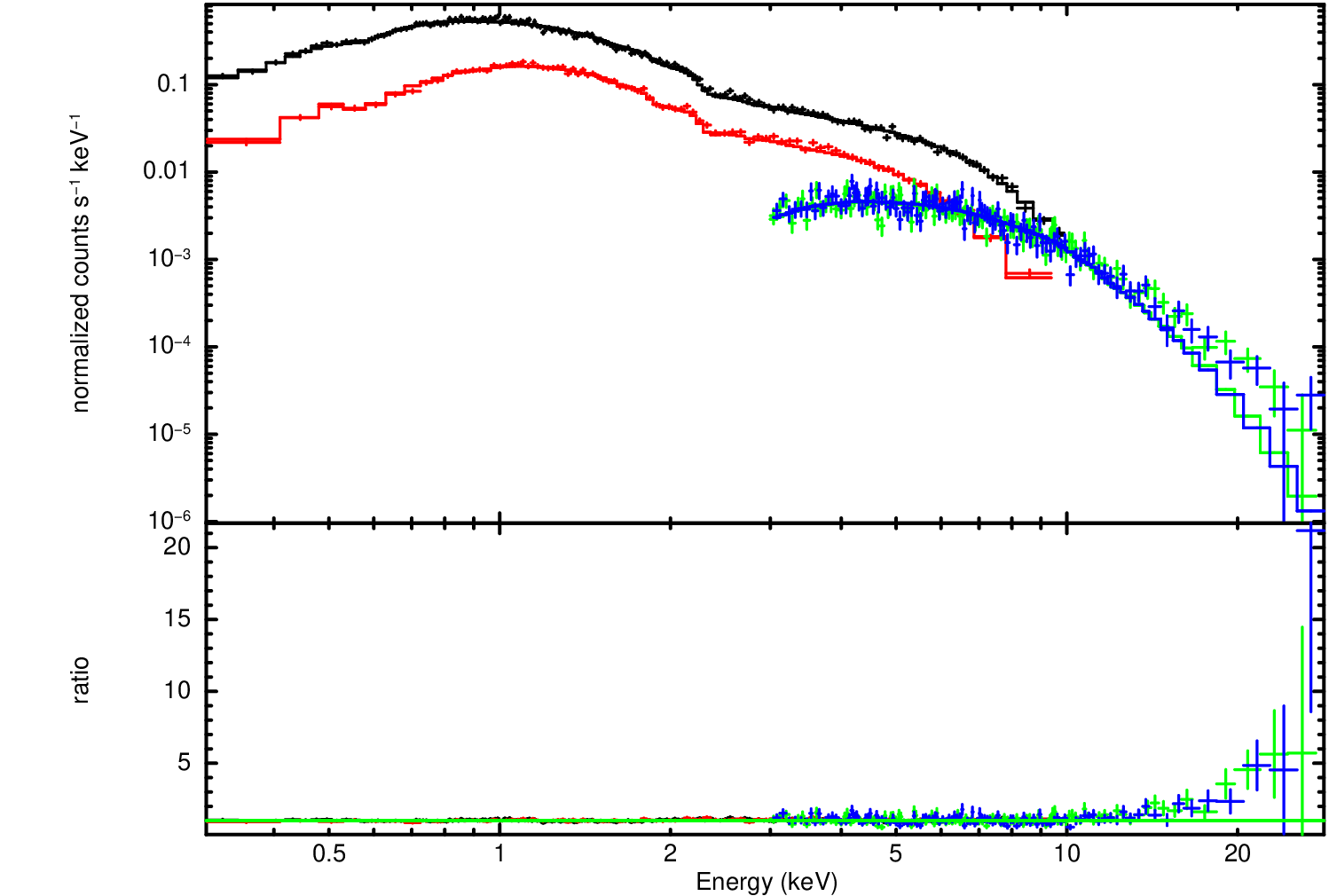}
\includegraphics[scale=0.35,angle=-90]{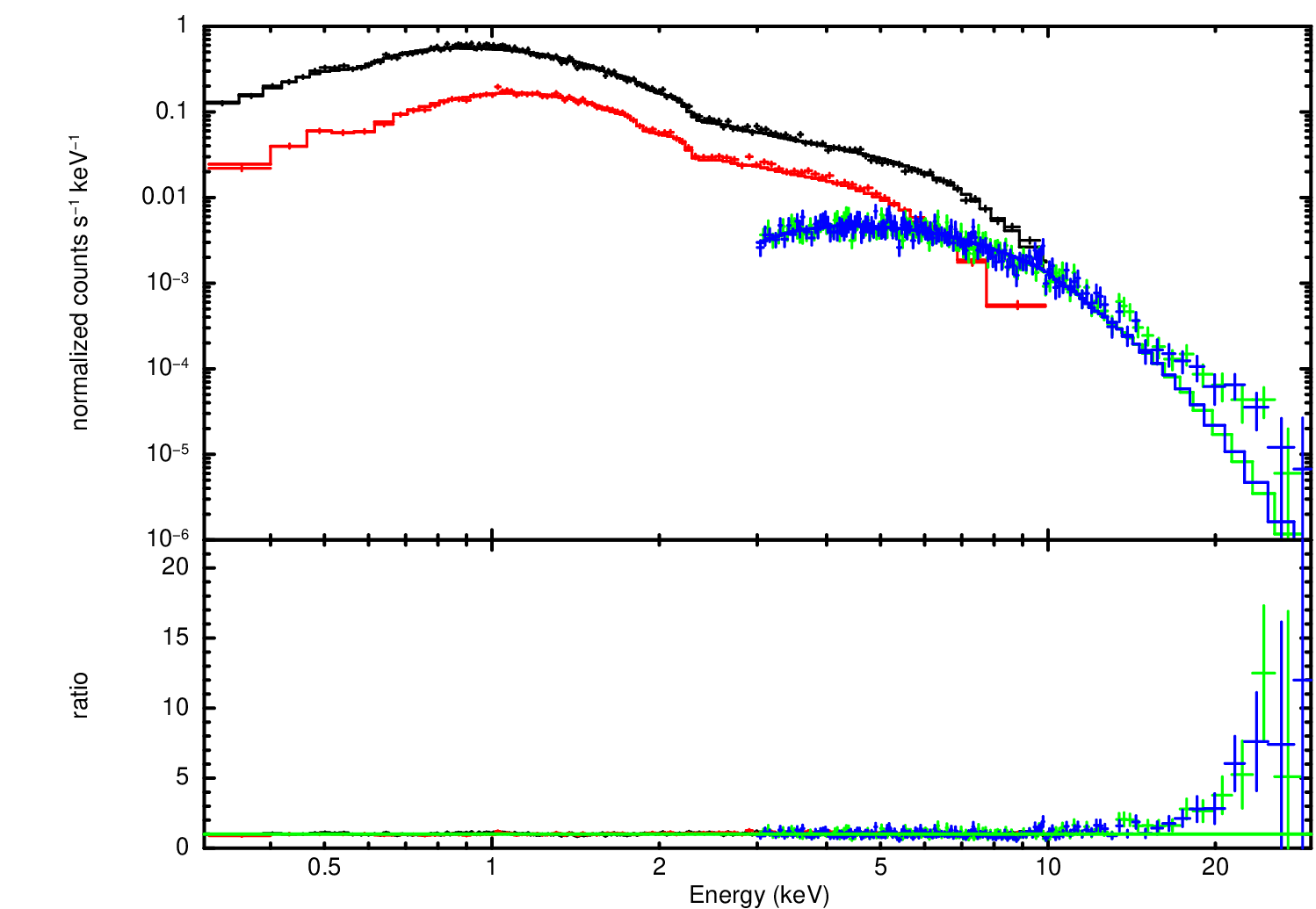}
\figcaption[t]{\footnotesize The {\it XMM-Newton} EPIC-pn (black),
  EPIC-MOS2 (red), and {\it NuSTAR} FPMA (green) and FPMB (blue)
  spectra of NGC 1313 X-1 are shown here.  Spectra from Epochs 1 and 2
  are shown in the left and right panels, respectively, with a toy model 
  for a ``patchy'' disk ($diskbb + diskbb$, see Table
  1).  This simple model provides a relatively good fit ($\chi^{2}/\nu
  \simeq 1.07$) in that it accounts for the bulk of the spectrum.
  However, it fails at high energy, likely due to unmodeled
  Comptonization.}
\end{figure}
\medskip

\clearpage

\begin{figure}
\begin{center}
\includegraphics[scale=0.35,angle=-90]{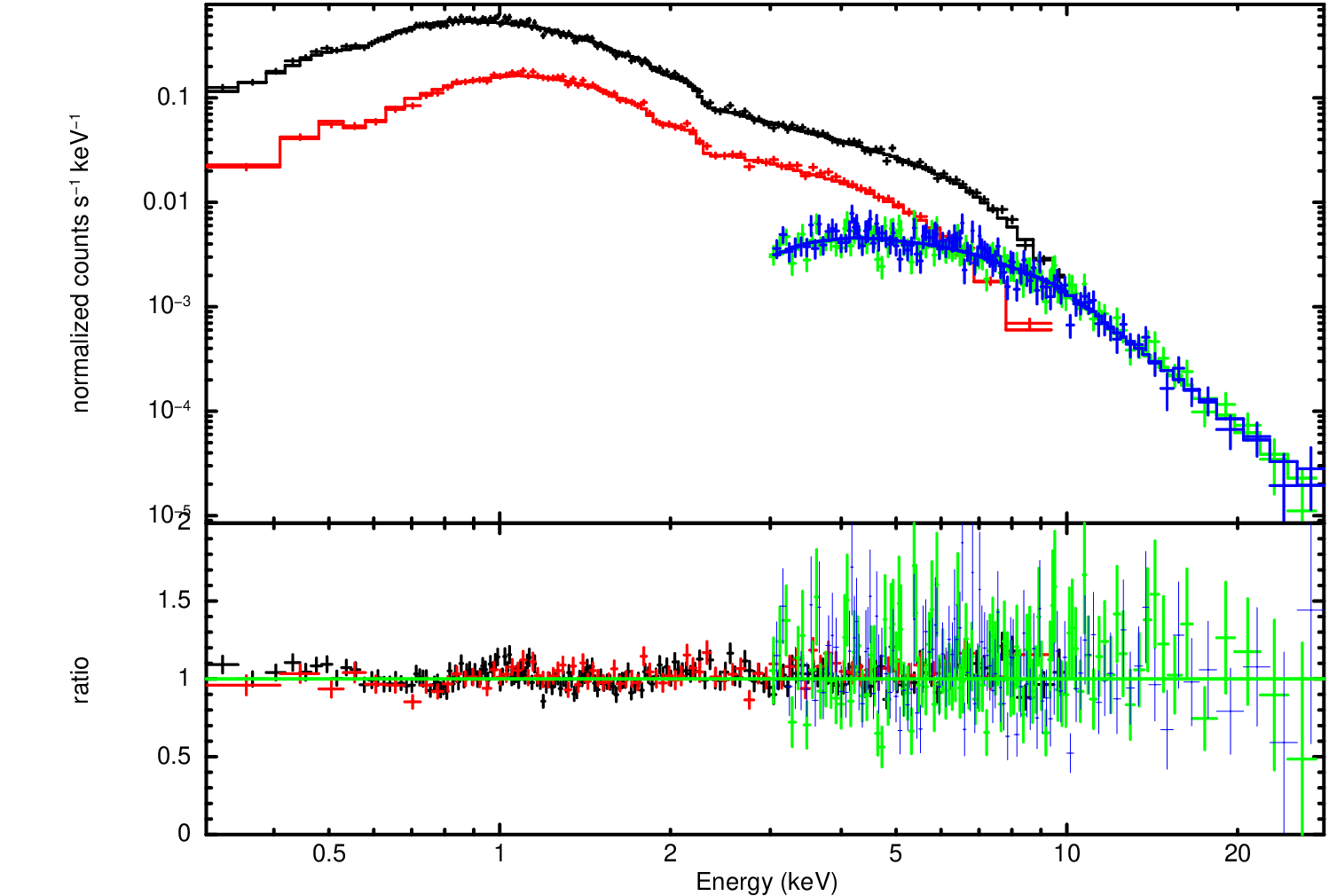}
\includegraphics[scale=0.35,angle=-90]{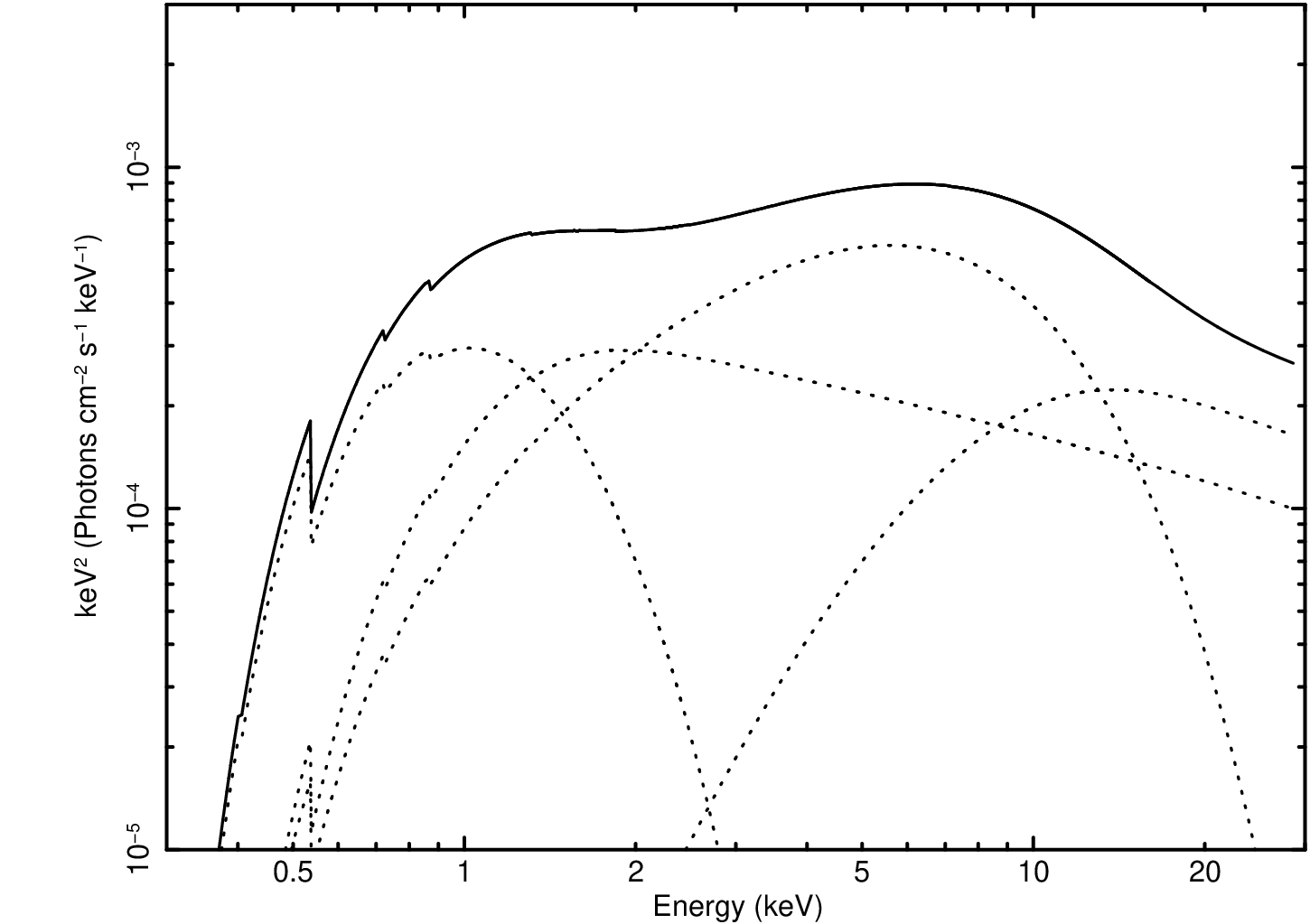}
\includegraphics[scale=0.35,angle=-90]{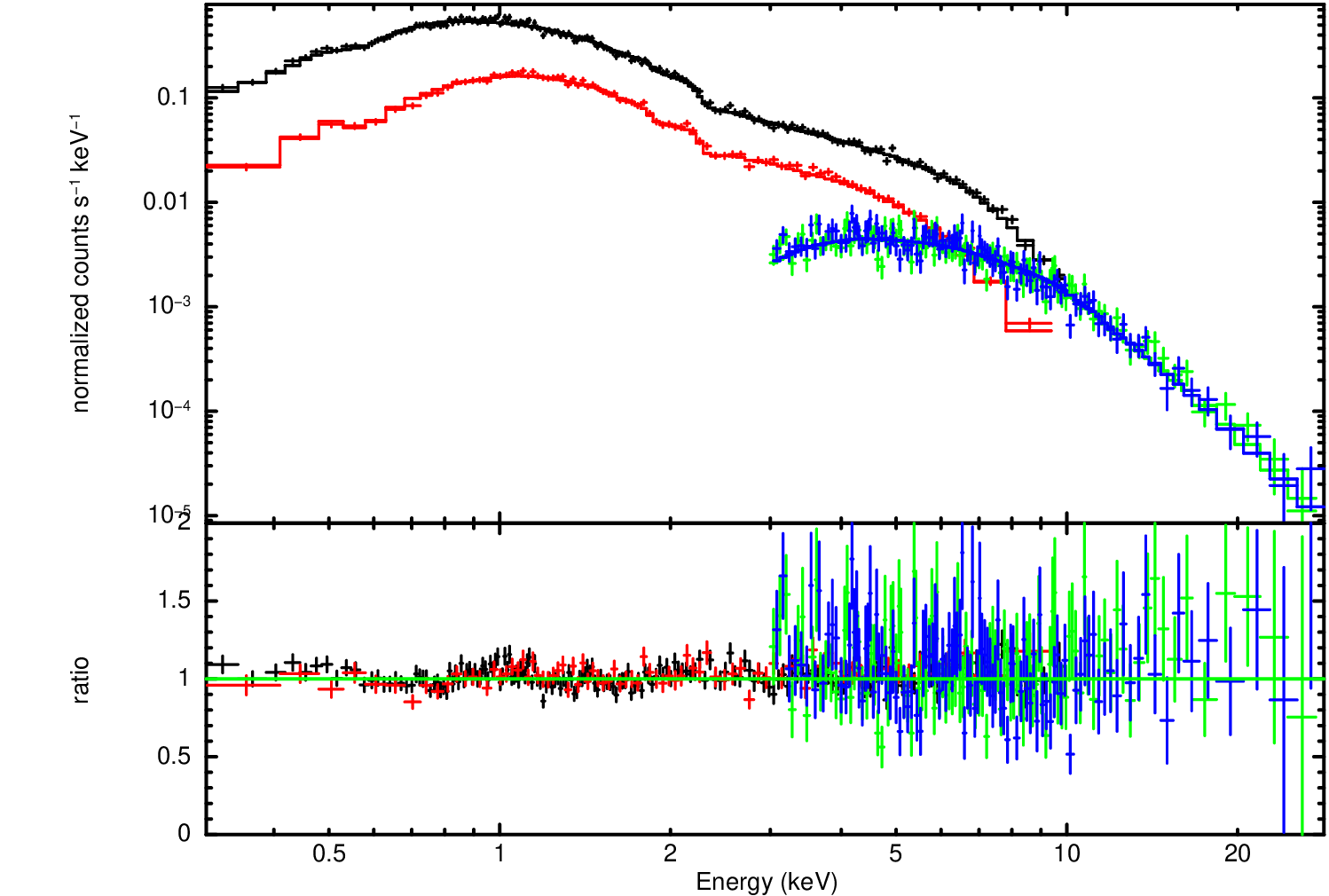}
\includegraphics[scale=0.35,angle=-90]{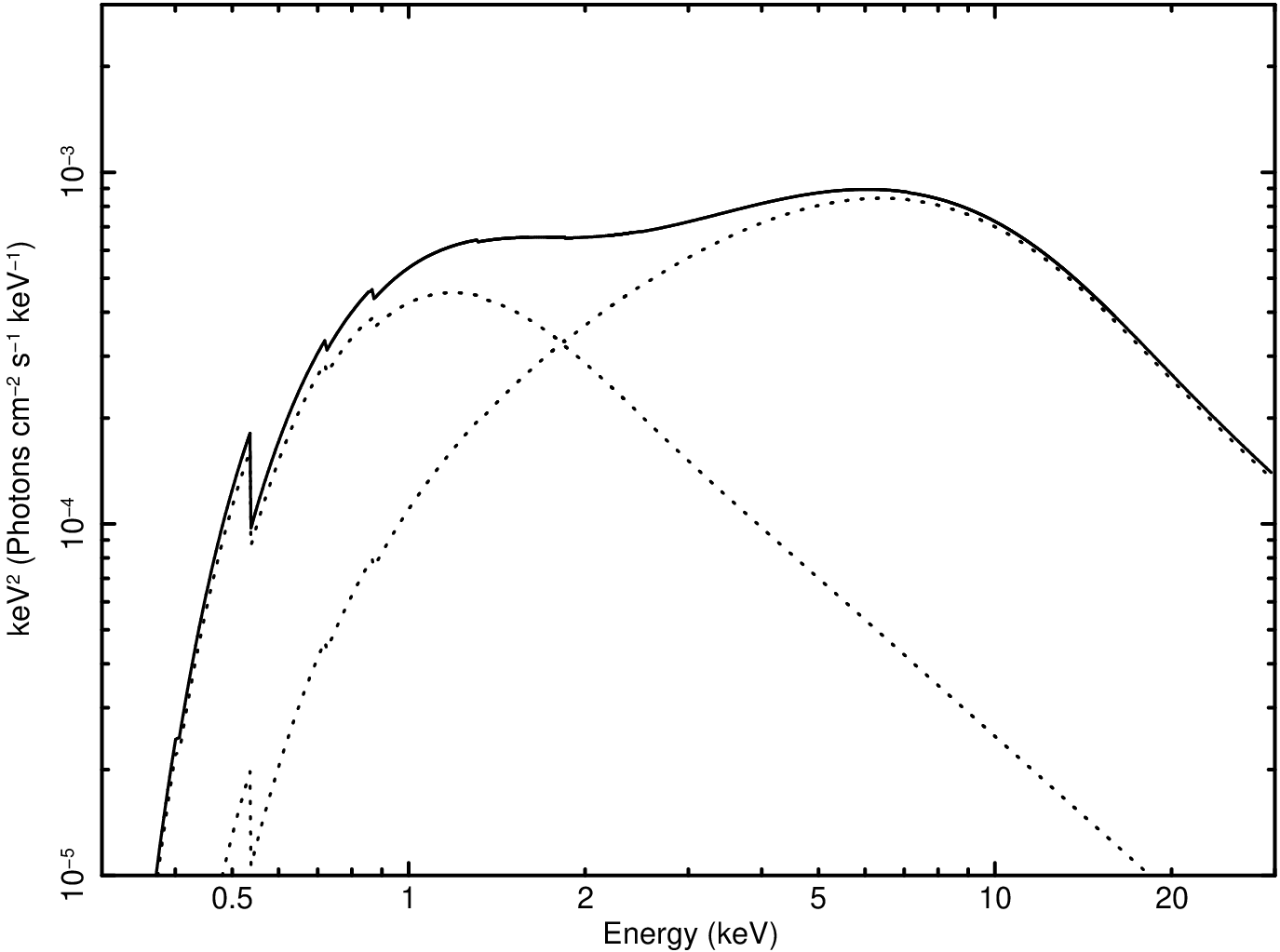}
\includegraphics[scale=0.35,angle=-90]{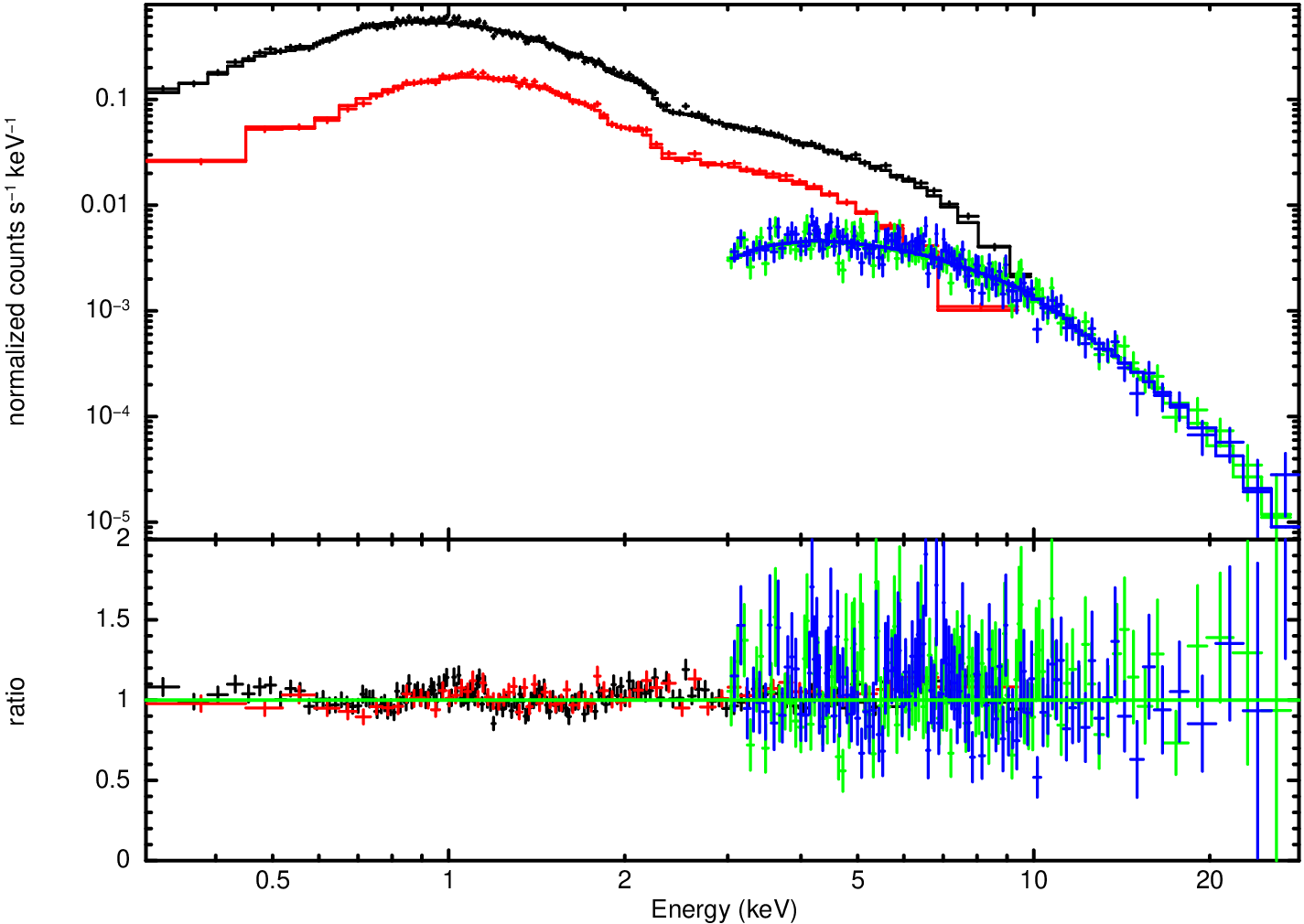}
\includegraphics[scale=0.35,angle=-90]{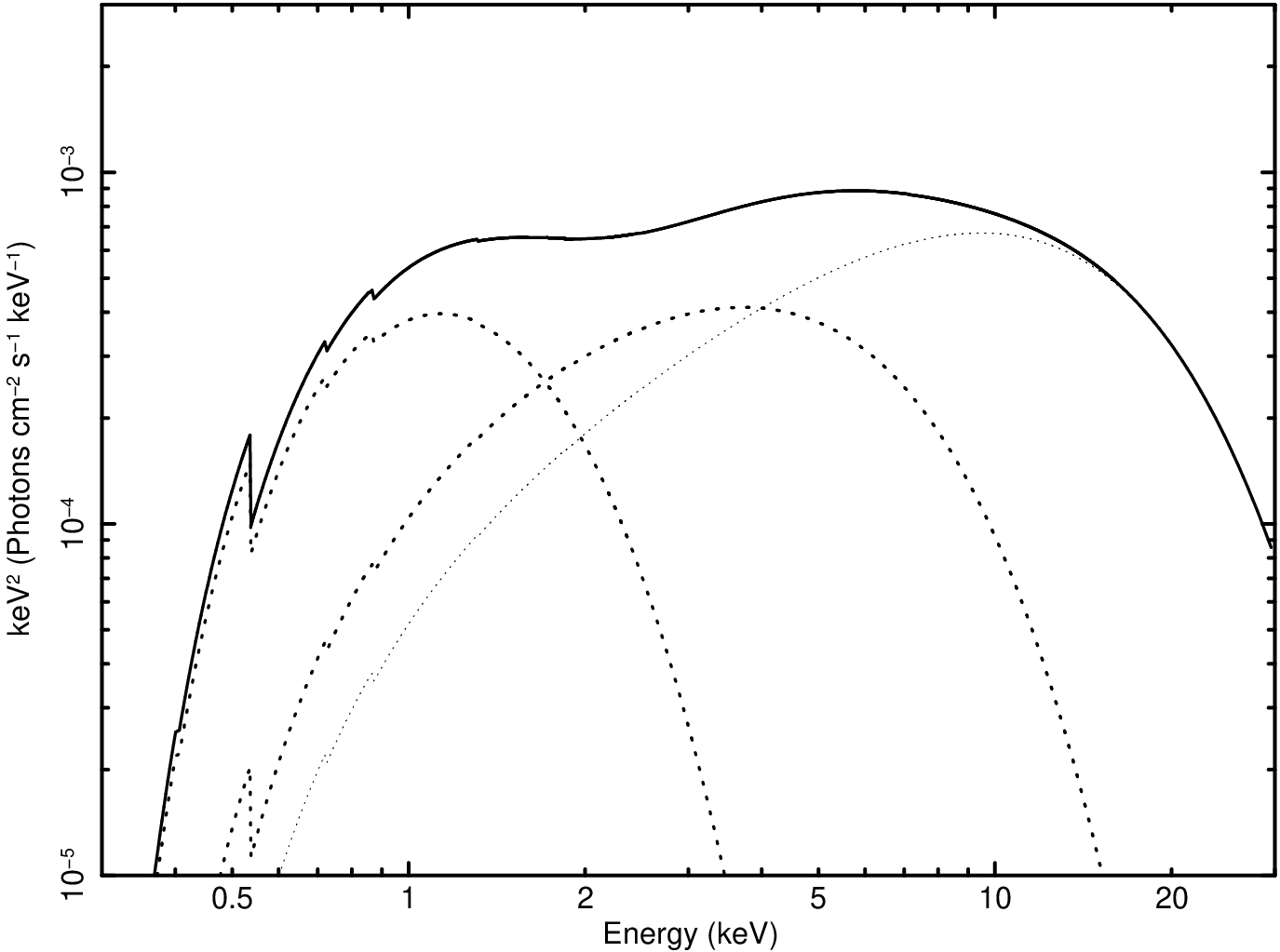}
\end{center}
\figcaption[t]{\footnotesize {\it XMM-Newton} EPIC-pn (black),
  EPIC-MOS2 (red), and {\it NuSTAR} FPMA (green) and FPMB (blue)
  spectra of NGC 1313 X-1 from Epoch 1 are shown here in the lefthand
  panels.  The righthand panels show the corresponding model, without
  the data.  In the top panels, two ``diskbb'' components were used to
  simulate a ``patchy'' disk, modified by Comptonization via
  ``compTT''.  In the middle panels, the ``compTT'' components were
  replaced by the ``simpl'' convolution model.  In the bottom panels,
  three disk blackbody components were fit, without Comptonization.
  Simple blackbody modela are also effective.  All of these simple
  models yield excellent fits (see Table 1).}
\end{figure}
\medskip

\clearpage

\begin{figure}
\includegraphics[scale=0.35,angle=-90]{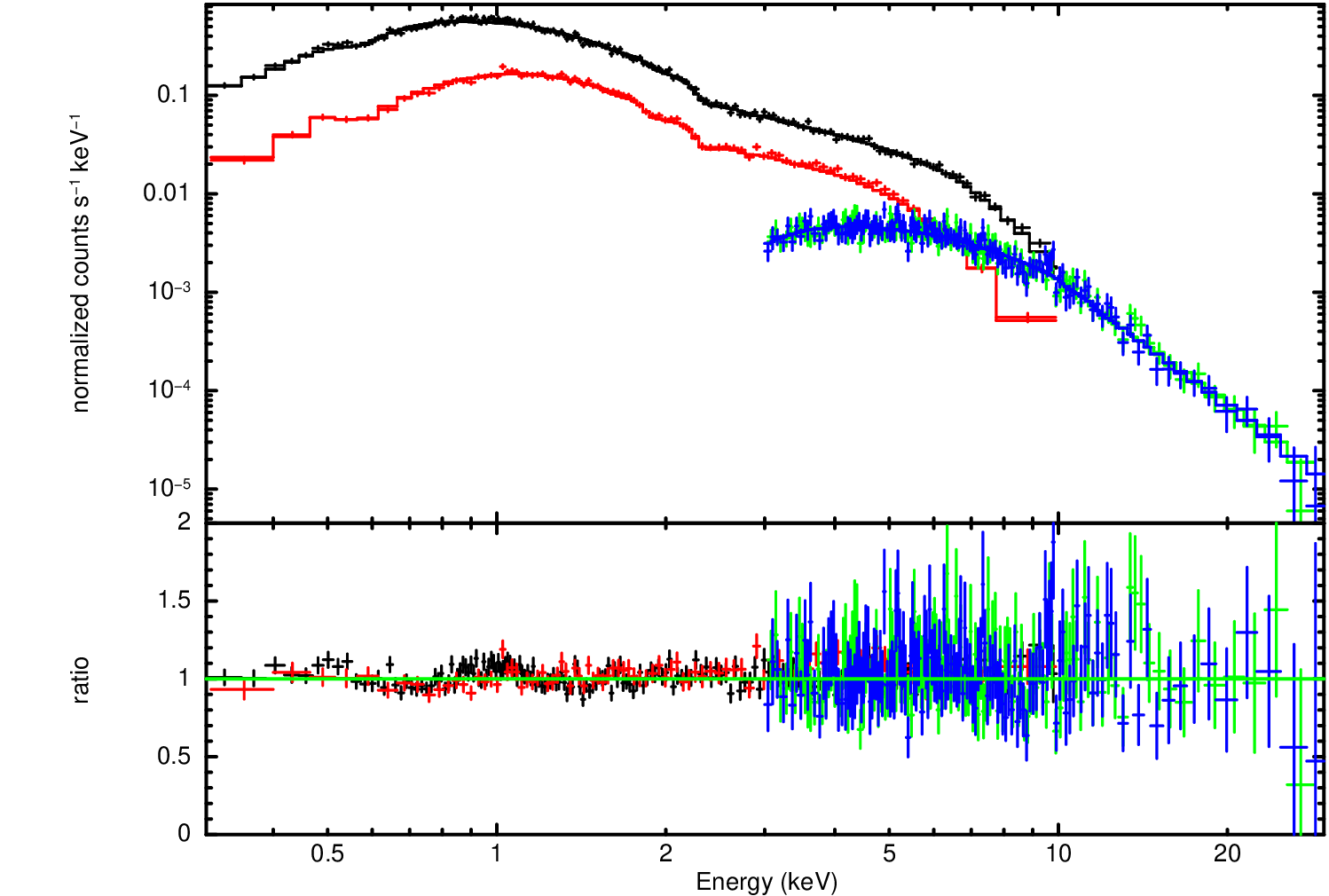}
\includegraphics[scale=0.35,angle=-90]{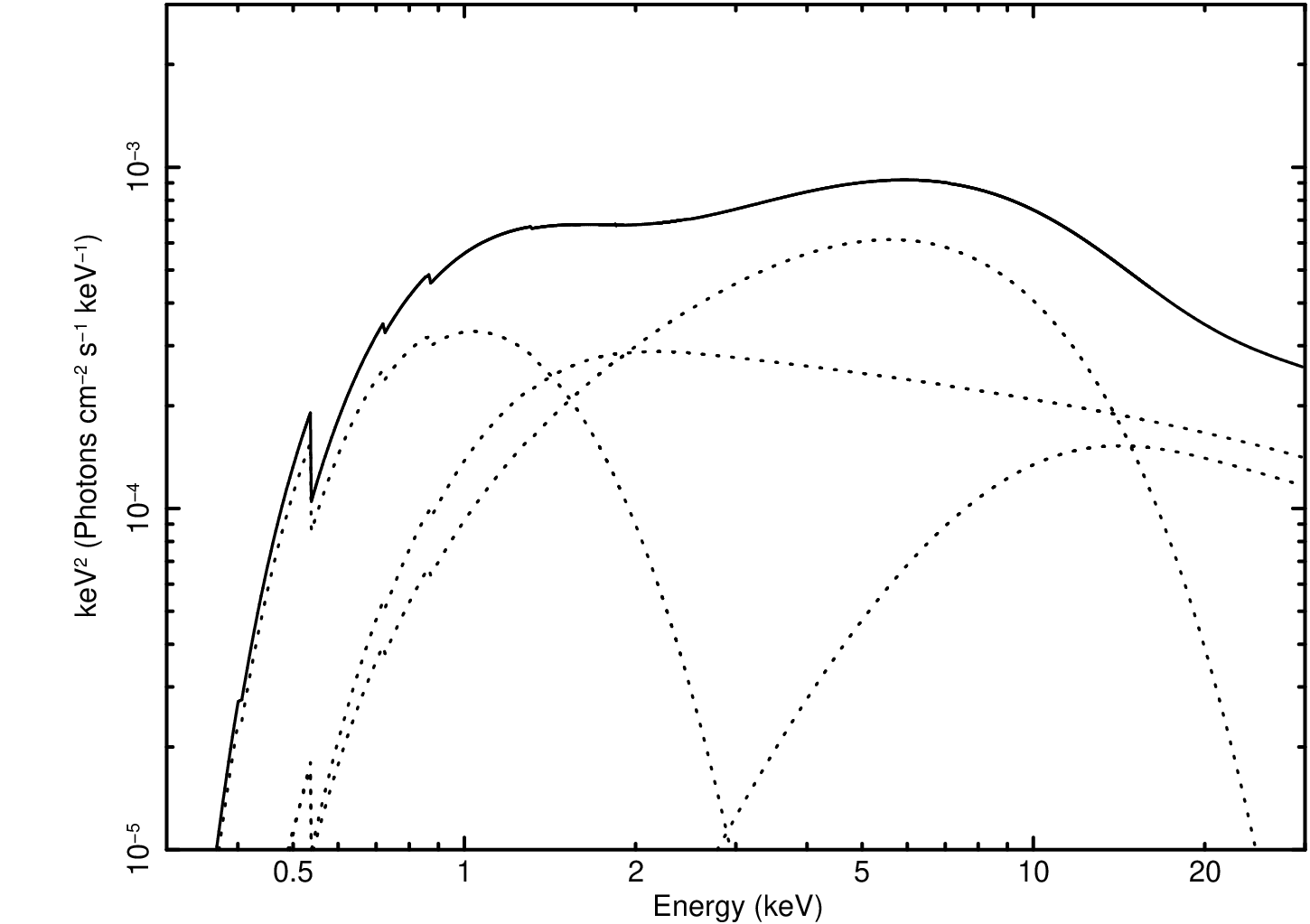}
\includegraphics[scale=0.35,angle=-90]{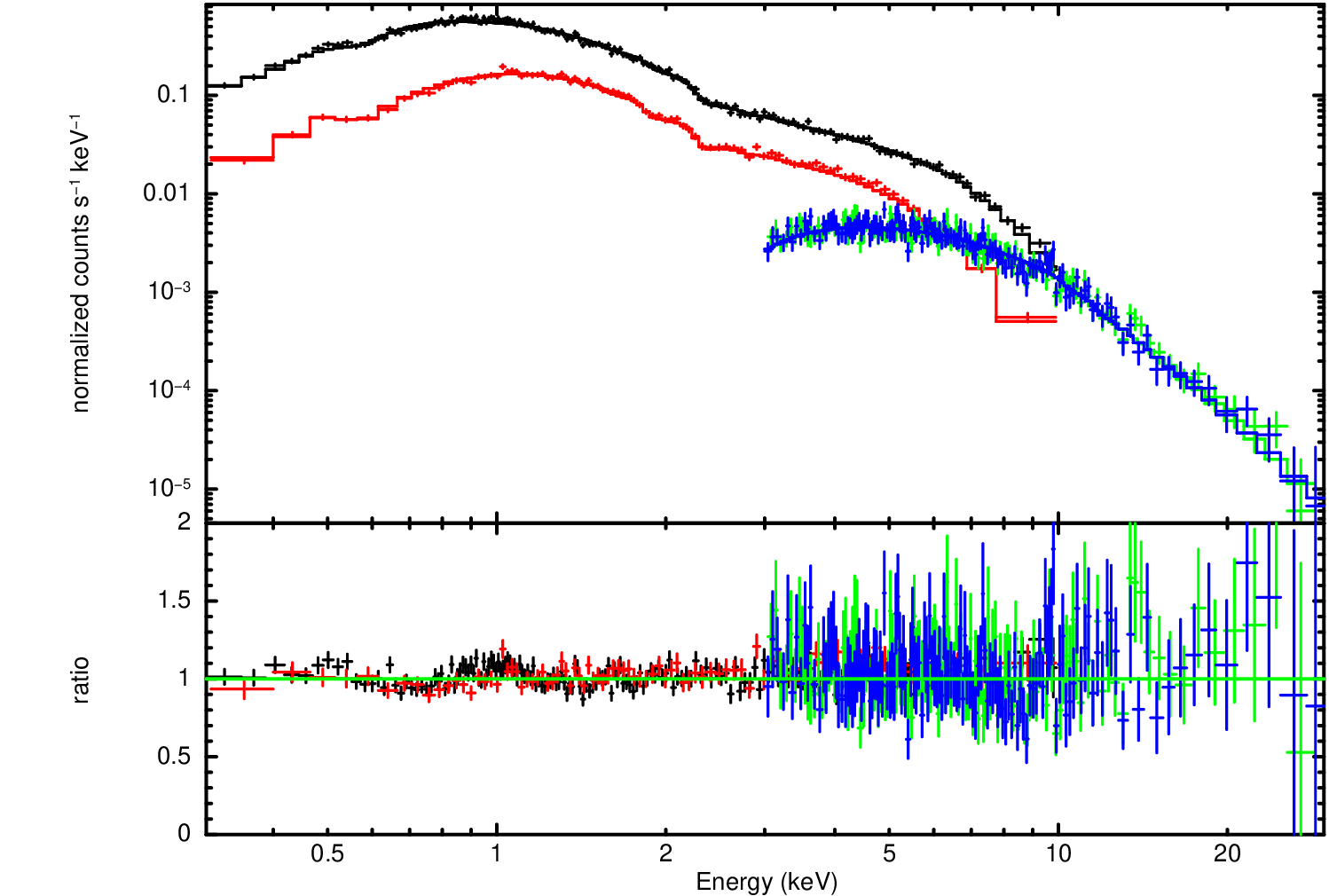}
\includegraphics[scale=0.35,angle=-90]{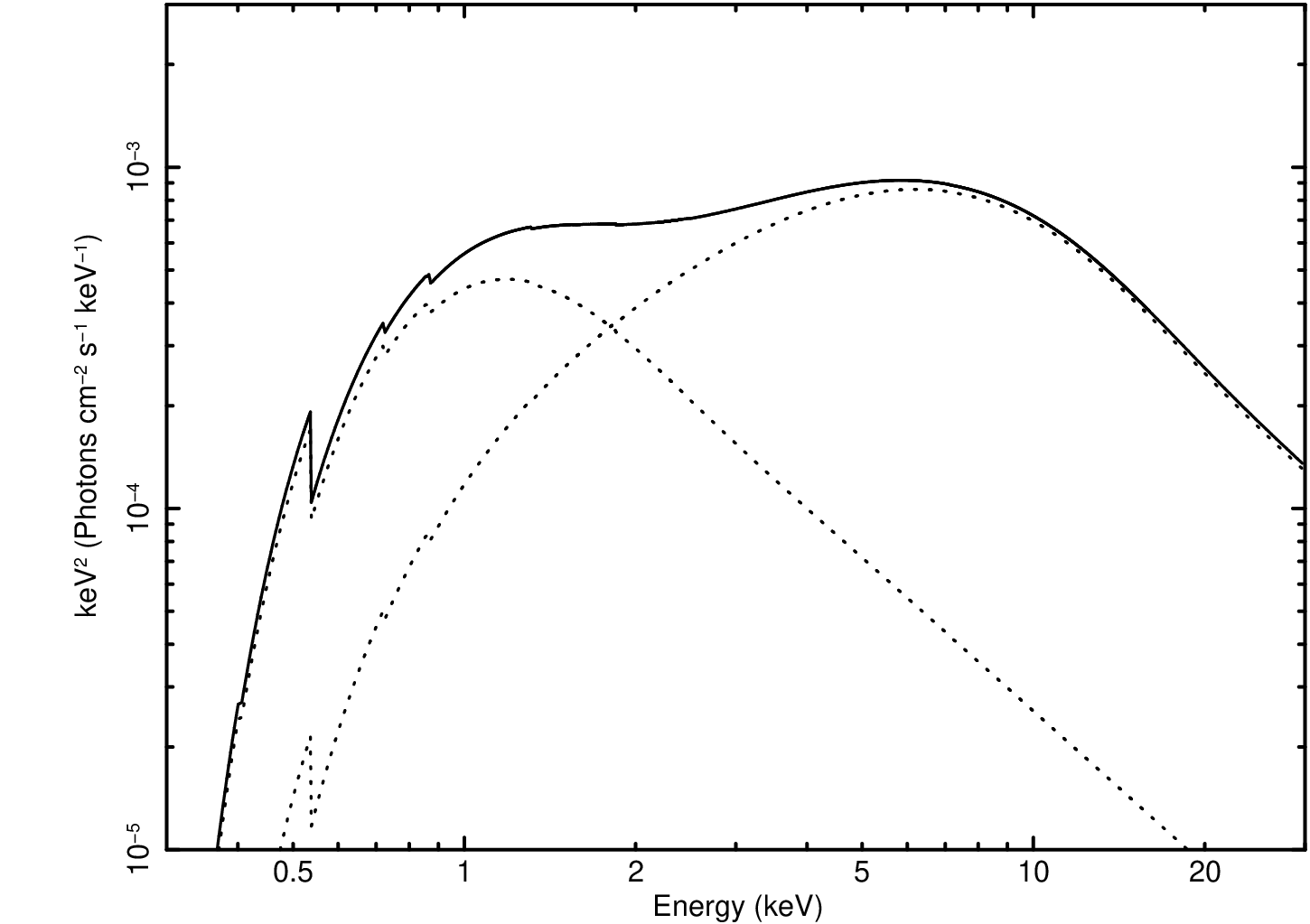}\\
\hspace{0.6in}\includegraphics[scale=0.35,angle=-90]{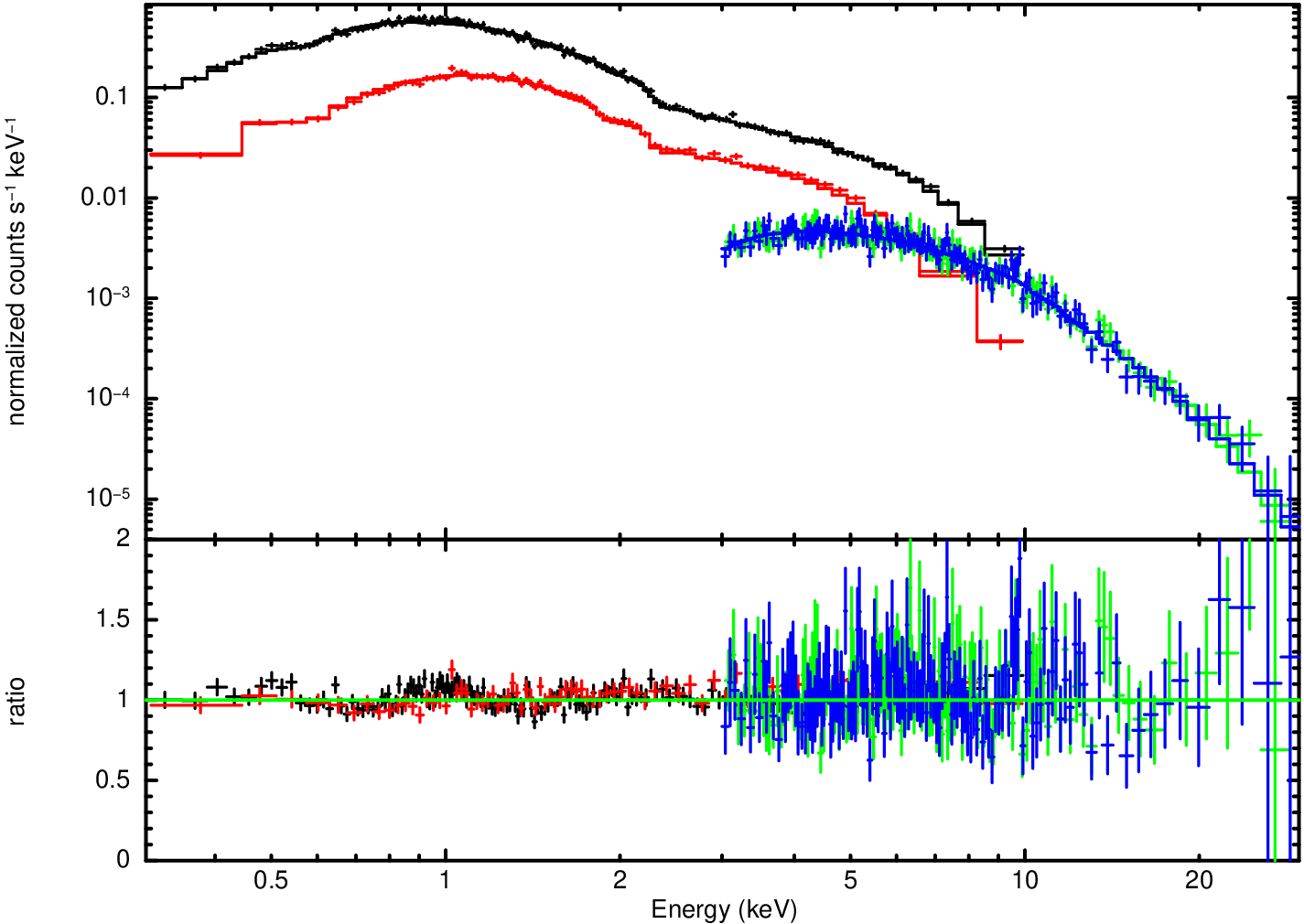}
\hspace{0.4in}\includegraphics[scale=0.35,angle=-90]{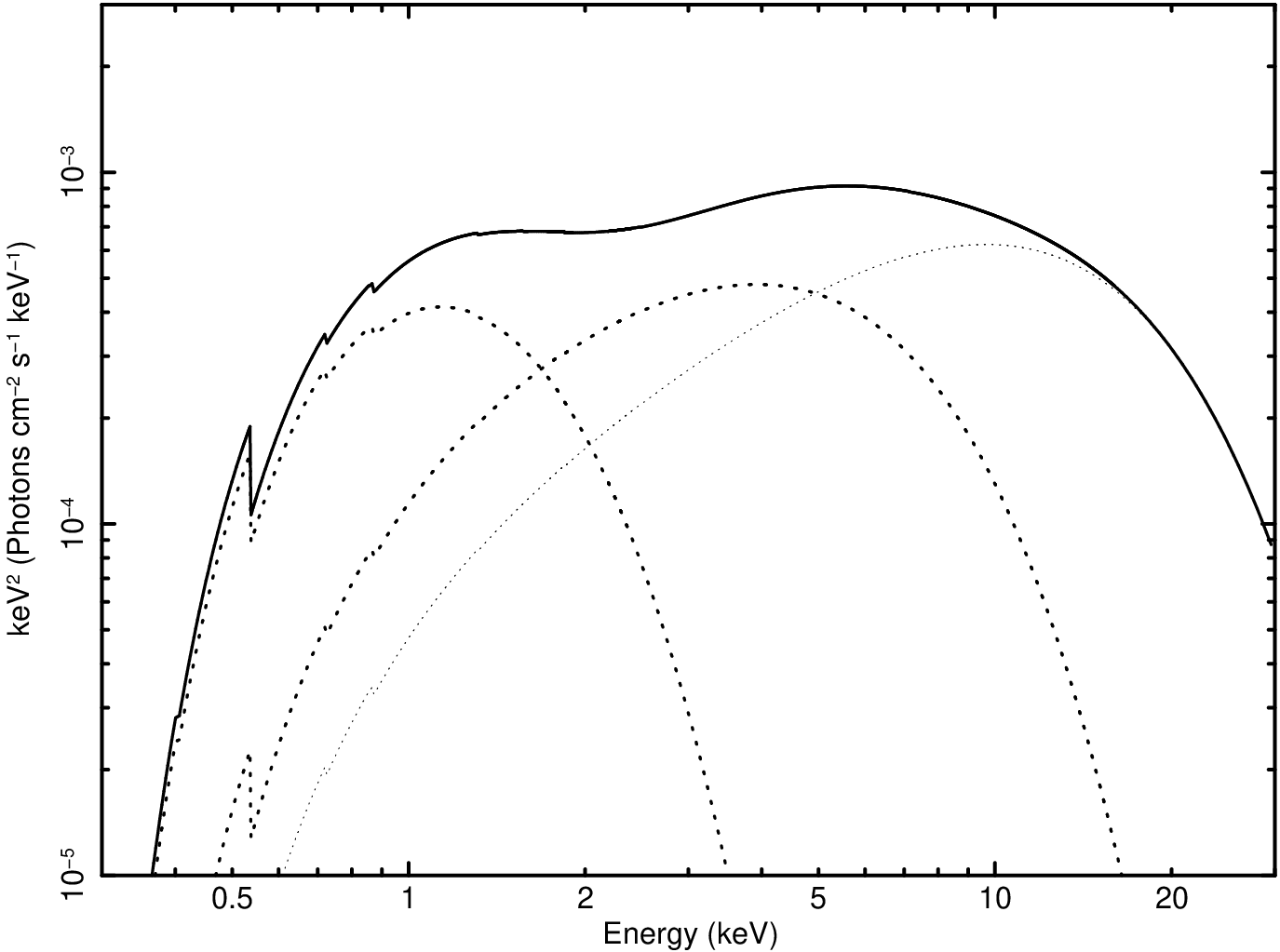}
\figcaption[t]{\footnotesize {\it XMM-Newton} EPIC-pn (black),
  EPIC-MOS2 (red), and {\it NuSTAR} FPMA (green) and FPMB (blue)
  spectra of NGC 1313 X-1 from Epoch 2 are shown here in the lefthand
  panels.  The righthand panels show the corresponding model, without
  the data.  In the top panels, two ``diskbb'' components were used to
  simulate a ``patchy'' disk, modified by Comptonization via
  ``compTT''.  In the middle panels, the ``compTT'' components were
  replaced by the ``simpl'' convolution model.  In the bottom panels,
  three disk blackbody components were fit, without Comptonization.
  Simple blackbody modela are also effective.  All of these simple
  models yield excellent fits (see Table 1).}
\end{figure}
\medskip



\begin{references}

\reference{} Armitage, P. J., \& Reynolds, C. S., 2003, MNRAS, 341, 1041

\reference{} Arnaud, K., 1996, Astronomical Data Analysis Software and
Systems V, ASP Converence Series, eds. G. H. Jacoby and J. Barnes,
101, 17

\reference{} Bachetti, M., et al., 2013, ApJ, 778, 163

\reference{} Begelman, M. C., 2001, ApJ, 551, 897

\reference{} Begelman, M. C., 2002, ApJ, 568, L97

\reference{} Cash, W., 1979, ApJ, 228, 939

\reference{} Cseh, D., Kaaret, P., Corbel, S., Grise, F., Lang, C.,
Koerding, E., Falcke, H., Jonker, P. G., Miller-Jones, J. C. A.,
Farrell, S., Yang, Y. J., Paragi, Z., \& Frey, S., 2014, MNRAS, in
press, arxiv:1311.4867

\reference{} Dexter, J., \& Quataert, E., 2012, MNRAS, 426, L71

\reference{} Dotan, C., \& Shaviv, N., 2011, MNRAS, 413, 1623

\reference{} Fabian, A. C., et al., 2009, Nature, 459, 540

\reference{} Gammie, C. F., 1998, MNRAS, 297, 929

\reference{} Gladstone, J., 2013, arxiv:1306.6886, in the proceedings of "X-ray Astronomy: towards the next 50 years", Milan Italy, 2012

\reference{} Gladstone, J., Roberts, T., Done, C., 2009, MNRAS, 397, 1836

\reference{} Greene, J. E., \& Ho, L. C., 2007, ApJ, 670, 92

\reference{} Grise, F., Kaaret, P., Corbel, S., Cseh, D., Feng, H.,
2013, MNRAS, 433, 1023

\reference{} Harrison, F. A., et al., 2013, ApJ, 770, 103

\reference{} Irwin, J. A., Athey, A. E., Bregman, J., 2003, ApJ, 587, 356

\reference{} Kajava, J., \& Poutanen, J., 2009, MNRAS, 398, 1450

\reference{} Kajava, J., Poutanen, J., Farrell, S., Grise, F., Kaaret,
P., 2012, MNRAS, 422, 990

\reference{} Kara, E., Fabian, A. C., Cackett, E. M., Miniutti, G.,
Uttley, P., 2013, MANRAS, 430, 1408

\reference{} King, A. R., Davies, M. B., Ward, M. J., Fabbiano, G.,
Elvis, M., 2001, ApJ, 552, L109

\reference{} Marshall, H., Canizares, C., Hillwig, T., Mioduszewski,
A., Rupen, M., Schulz, N., Nowak, M., Heinz, S., 2013, ApJ, 775, 75

\reference{} McClintock, J. E., \& Remillard, R., 2006, in ``Compact
Stellar X-ray Sources'', eds. W. H. G. Lewin \& M. van der Klis,
Cambridge: Cambridge Univesity Press

\reference{} Merloni, A., \& Fabian, A. C., 2001, MNRAS, 321, 549

\reference{} Merloni, A., Fabian, A. C., \& Ross, R. R., 2000, MNRAS, 313, 193

\reference{} Middleton, M., Sutton, A., \& Roberts, T., 2011, MNRAS, 417, 464

\reference{} Middleton, M., Walton, D., Roberts, T., Heil, L., 2014,
MNRAs, 438, L51

\reference{} Miller, J. M., Fabbiano, G., Miller, M. C., Fabian,
A. C., 2003, ApJ, 585, L37

\reference{} Miller, J. M., Fabian, A. C., \& Miller, M. C., 2004, ApJ, 614, L117

\reference{} Miller, J. M., Walton, D. J., King, A. L., Reynolds,
M. T., Fabian, A. c., Miller, M. C., Reis, R. C., 2013, ApJ, 776, L36

\reference{} Mitsuda, K., Inoue, H., Koyama, K., Makishima, K.,
Matsuoka, M., Ogawara, Y., Suzuki, K., Tanaka, Y., Shibazaki, N.,
Hirano, T., 1984, PASJ, 37, 741

\reference{} Pasham, D., \& Strohmayer, T. E., 2012, ApJ, 753, 139

\reference{} Pasham, D., \& Strohmayer, T. E., 2013, ApJ, 764, 93

\reference{} Pintore, F., \& Zampieri, L., 2012, MNRAS, 420, 1107

\reference{} Rana, V., et al., 2014, ApJ, submitted

\reference{} Shakura, N., \& Sunyaev, R., 1973, A\&A, 24, 337

\reference{} Soria, R., 2011, Astronomische Nachrichten, 332, 330

\reference{} Steiner, J., Narayan, R., McClintock, J., Ebisawa, K.,
2009, PASP, 885, 1279

\reference{} Strohmayer, T. E., \& Mushotzky, R. F., 2003, ApJ, 586, L61

\reference{} Strohmayer, T. E., Mushotzky, R. F., Winter, L., Soria,
R., Uttley, P., Cropper, M., 2007, ApJ, 660, 580

\reference{} Sutton, A. D., Roberts, T. P., Walton, D. J., Galdstone,
J. C., Scott, A. E., 2012, MNRAS, 423, 1154

\reference{} Titarchuk, L., 1994, ApJ, 434, 570

\reference{} Tomsick, J. A., Kaaret, P., Kroeger, R., Remillard, R.,
1999, ApJ, 512, 892

\reference{} Tully, R. B., 1988, Nearby Galaxies Catalog (Cambridge:
Cambridge University Press)

\reference{} Walton, D. J., Miller, J. M., Reis, R., C., Fabian,
R. C., 2012, MNRAS, 426, 473

\reference{} Walton, D. J., et al., 2013a, ApJ, 779, 148

\reference{} Walton, D. J., Miller, J. M., Harrison, F. A., Fabian,
A. C., Roberts, T. P., Middleton, M. J., Reis, R. C., 2013b, ApJ, 773,
L9

\reference{} Walton, D. J., et al., 2014, ApJ, submitted

\reference{} Wilms, J., Allen, A., McCray, R., 2000, ApJ, 542, 914




\end{references}
\end{document}